\documentstyle[]{mn}
\newif\ifAMStwofonts

\def\etal{{\rm et al.}}
\def\cm{{\rm cm}}
\def\kms{{\rm\,km\,s^{-1}}}
\def\simgt{\mathrel{\spose{\lower 3pt\hbox{$\sim$}}
        \raise 2.0pt\hbox{$>$}}}
\def\simlt{\mathrel{\spose{\lower 3pt\hbox{$\sim$}}
        \raise 2.0pt\hbox{$<$}}}

\ifoldfss
  \ifCUPmtlplainloaded \else
    \NewTextAlphabet{textbfit} {cmbxti10} {}
    \NewTextAlphabet{textbfss} {cmssbx10} {}
    \NewMathAlphabet{mathbfit} {cmbxti10} {} 
    \NewMathAlphabet{mathbfss} {cmssbx10} {} 
  \fi
  \ifAMStwofonts
    \ifCUPmtlplainloaded \else
      \NewSymbolFont{upmath} {eurm10}
      \NewSymbolFont{AMSa} {msam10}
      \NewMathSymbol{\upi}     {0}{upmath}{19}
      \NewMathSymbol{\umu}     {0}{upmath}{16}
      \NewMathSymbol{\upartial}{0}{upmath}{40}
      \NewMathSymbol{\leqslant}{3}{AMSa}{36}
      \NewMathSymbol{\geqslant}{3}{AMSa}{3E}

    \fi
  \fi
\fi 

\ifnfssone
  \newmathalphabet{\mathit}
  \addtoversion{normal}{\mathit}{cmr}{m}{it}
  \addtoversion{bold}{\mathit}{cmr}{bx}{it}
  \newmathalphabet{\mathbfit} 
  \addtoversion{normal}{\mathbfit}{cmr}{bx}{it}
  \addtoversion{bold}{\mathbfit}{cmr}{bx}{it}
  \newmathalphabet{\mathbfss} 
  \addtoversion{normal}{\mathbfss}{cmss}{bx}{n}
  \addtoversion{bold}{\mathbfss}{cmss}{bx}{n}
  \ifAMStwofonts
    \ifCUPmtlplainloaded \else
      \UseAMStwoboldmath
      \makeatletter
      \new@mathgroup\upmath@group
      \define@mathgroup\mv@normal\upmath@group{eur}{m}{n}
      \define@mathgroup\mv@bold\upmath@group{eur}{b}{n}
      \edef\UPM{\hexnumber\upmath@group}
      \new@mathgroup\amsa@group
      \define@mathgroup\mv@normal\amsa@group{msa}{m}{n}
      \define@mathgroup\mv@bold\amsa@group{msa}{m}{n}
      \edef\AMSa{\hexnumber\amsa@group}
      \makeatother
      \mathchardef\upi="0\UPM19
      \mathchardef\umu="0\UPM16
      \mathchardef\upartial="0\UPM40
      \mathchardef\leqslant="3\AMSa36
      \mathchardef\geqslant="3\AMSa3E
    \fi
  \fi
\fi 

\ifnfsstwo
  \DeclareMathAlphabet{\mathbfit}{OT1}{cmr}{bx}{it}
  \SetMathAlphabet\mathbfit{bold}{OT1}{cmr}{bx}{it}
  \DeclareMathAlphabet{\mathbfss}{OT1}{cmss}{bx}{n}
  \SetMathAlphabet\mathbfss{bold}{OT1}{cmss}{bx}{n}
  \ifAMStwofonts
    \ifCUPmtlplainloaded \else
      \DeclareSymbolFont{UPM}{U}{eur}{m}{n}
      \SetSymbolFont{UPM}{bold}{U}{eur}{b}{n}
      \DeclareSymbolFont{AMSa}{U}{msa}{m}{n}
      \DeclareMathSymbol{\upi}{0}{UPM}{"19}
      \DeclareMathSymbol{\umu}{0}{UPM}{"16}
      \DeclareMathSymbol{\upartial}{0}{UPM}{"40}
      \DeclareMathSymbol{\leqslant}{3}{AMSa}{"36}
      \DeclareMathSymbol{\geqslant}{3}{AMSa}{"3E}
    \fi
  \fi
\fi 

\ifCUPmtlplainloaded \else
  \ifAMStwofonts \else 
    \def\upi{\pi}
    \def\umu{\mu}
    \def\upartial{\partial}
  \fi
\fi

\title[Application of the Contouring Method to Extended Microlensed Sources]
  {Application of the Contouring Method to Extended Microlensed Sources}
\author[J. S. B. Wyithe \& R. L. Webster]
  {J.~ S.~ B.~Wyithe,$^1$ 
  R.~L.~Webster,$^1$\\
  $^1$ School of Physics, University of Melbourne, Parkville, Vic, 3052, 
Australia\\
 Email: swyithe@physics.unimelb.edu.au, rwebster@physics.unimelb.edu.au }
\date{Accepted  Received }
\pagerange{\pageref{firstpage}--\pageref{lastpage}}
\pubyear{1998}

\def\LaTeX{L\kern-.36em\raise.3ex\hbox{a}\kern-.15em
    T\kern-.1667em\lower.7ex\hbox{E}\kern-.125emX}

\begin{document}
\label{firstpage}

\maketitle

\begin{abstract}

The method devised by Lewis et al. (1993) for calculating the light curve of a microlensed point source is expanded to two dimensions to enable the calculation of light curves of extended sources. This method is significantly faster than the ray shooting method that has been used in the past. The increased efficiency is used to obtain much higher resolution light curves over increased time scales. We investigate the signatures arising from different source geometries in a realistic microlensing model. We show that a large fraction of high magnification events (HMEs) in image A of Q2237+0305 involve only one caustic, and could therefore yield information on the structure of the quasar continuum through the recognition of a characteristic event shape. In addition, the cataloguing of HMEs into morphological type will, in theory, enable the direction of the transverse motion, as well as the source size to be obtained from long term monitoring.

\end{abstract}


\section{Introduction}

 Much of the study of cosmological microlensing has focussed on the modelling of the images of Q2237+0305 in which microlensing was first observed by Irwin et al. (1989) and Corrigan et al. (1991). Many authors (e.g. Paczynski 1986; Wambsganss \& Paczynski 1991; Witt, Kayser \& Refsdal 1993; Lewis et al. 1996) have used different numerical methods to model variations in the observed flux of this object.  Statistical arguments 
have then been used to place constraints on the size of the emission regions of the quasar.  The simulations have also shown how long term monitoring of
microlensing events can yield information on the composition of the lensing galaxy. As an example, Schmidt \& Wambsganss (1998) place limits on the lower end of the mass range from an observed lack of microlensing in Q0957+561.

Interpretation of model light curves for extended sources calculated using ray shooting (e.g. Kayser, Refsdal \& Stabell 1986; Wambsganss \& Paczynski 1991) has been hindered by the poor statistical coverage of the simulated light curves.  The majority of magnification maps have been composed of 500$\times$500 or 1000$\times$1000 pixels. Therefore when the resolution is high enough, the light curves have only covered a few months or years for source sizes of $<10^{16}cm$, instead of the centuries required to adequately sample high magnification events.
In addition to the ray-shooting method, approximations to microlensed light curves for extended sources
have been obtained in two ways: through combining the parametric representation of caustics with the ray-shooting method (Wambsganss, Witt \& Schneider 1992), as well as through integrating an extended profile over a line of point source magnifications (Witt \& Mao 1994). The first of these methods computes the light curve at a caustic by convolving the near caustic approximation (Eqn \ref{nearcaust}) with the source intensity profile. It therefore assumes the validity of Eqn \ref{nearcaust} for a complex caustic network. The second approach does not consider the behaviour of the magnification pattern in the direction perpendicular to the source line, and will thus overestimate the proportion of HMEs which involve only one caustic.

 High magnification events
can be classified into a number of different classes.  If the event is
produced by the source crossing a single caustic (SHME), then the characteristic
shape of the caustic will contain information on the geometry of the
quasar emission regions (Fluke \& Webster 1998). On the other hand,
high magnification events produced by the source crossing two or more
fold caustics (MHMEs) are harder to interpret. Classification
of HMEs, and the subsequent physical interpretation, requires full modelling
of individual events. This places limits on the coarseness of the microlensing simulations that may be of use.  The observed microlensed event in image A of Q2237+0305 in 1990 had a rise time $\sim$ 26 days. Therefore as a rough guide, a temporal resolution of $\sim 1 $ day is required for the successful modelling of a well sampled HME in Q2237+0305. 
Lewis et al. (1993) and Witt (1993) independently introduced a method for calculating the images of a line, and therefore the light curve of a point source, microlensed by a collection of point masses. Section \ref{Contouring} describes how this method can
be expanded to calculate the microlensed light curve of an extended source with an arbitrary surface brightness profile.  In section \ref{Simulations} this extended method is used to compute microlensed light curves of simple model quasars with different intensity profiles. The number and type of high magnification events are then determined.
Throughout the paper, standard notation for gravitational microlensing is used.
The Einstein radius of a 1M$_{\odot}$ star in the lens plane is denoted by $\xi_{o}$; in the source plane it is $\eta_{o}$; the normalised shear due to mass in the galaxy that is outside the region important for microlensing is $\gamma$ and the convergence or optical depth is $\kappa$. Where required, a cosmology having $\Omega=1$ and $\Lambda=0$ with $H_{o}=75\,km\,sec^{-1}\,Mpc^{-1}$ is assumed.

\section{Contouring Algorithm for Extended Sources}
\label{Contouring}
\subsection{Construction Of The Light Curve For An Extended Source}

The method used to compute the light curve of an extended source is
based on the contouring method.
A series of microlensed light curves are computed. These light curves have parallel source tracks that together 
cover the source and produce a regular grid of points in the source plane. A source intensity profile is then be convolved with this grid to produce a total observed
magnification of the source. When an extended source is microlensed, its magnification at the point $\vec{\eta}$ is calculated using the convolution:

\begin{equation}
\label{2dint}
\mu(\vec{\eta}\,)=\frac{\int P(\vec{\eta}\,'-\vec{\eta}\,)\mu_{p}(\vec{\eta}\,'\,)d^{2}\eta'}{\int P(\vec{\eta}\,'-\vec{\eta}\,)d^{2}\eta'},
\end{equation}
where $P(\vec{\eta}\,'-\vec{\eta})$ is the intensity profile of the source centred on $\vec{\eta}$ at the position $\vec{\eta}\,'$, and $\mu_{p}(\vec{\eta})$ is the 2-dimensional magnification pattern. If the magnification pattern is calculated using the ray shooting technique, then the computed magnification at the coordinates of a pixel is the average magnification over the region of source plane covered by that pixel.  Thus the convolution (\ref{2dint}) can be easily evaluated numerically as a weighted average. However, if the contouring method is used to build a magnification grid in the source plane, then the magnification at a given grid point is that of a point source. The singularities in $\mu_{p}(\vec{\eta})$ at 
\begin{equation}
\{\vec{\eta}\,\}=\{\vec{\eta}\,(\vec{\xi}):\frac{1}{\mu_{p}(\vec{\eta}\,(\vec{\xi}))}=0\},
\end{equation}
 where $\vec{\xi}$ is the image position, must be integrated out directly as the magnification of the point source is not necessarily a representative average of the region surrounding that point. This is the most important aspect of the calculation as much of the flux and therefore the information in the light curve originates from the parts of the source which lie very close to a caustic.

To calculate the two dimensional integral (Eqn \ref{2dint}) over the finite grid, the $1$-dimensional integrals for each line of the grid are first evaluated. The two dimensional integral is then evaluated by integrating these areas perpendicularly to the direction of the source line. 
The number of lines that cover the source strip was determined as follows. The transverse velocity is unknown, however as the event rise time was measured at $\sim$ 26 days, we know that $\sim$ 50 points per source width will sample to the same level as potential observation. We therefore set 50 lines as the minimum coverage for the strip.

 The convolution of the source intensity profile with each source line is performed in sections. A source which lies on the side of a caustic line such that it has two associated critical images is referred to as being inside the caustic. The parts of the line that do not lie just inside a caustic are smooth and so can be integrated in a straight forward manner. For sections immediately inside a caustic the magnification is highly nonlinear with distance. To overcome this nonlinearity we compute the numerical integration across the caustics with the help of the analytical result of Chang \& Refsdal (1979) and Chang (1984).
The magnification $\mu_{p}$ of a point source at a small perpendicular distance $d$ from a fold caustic is

\begin{equation}
\label{nearcaust}
\mu_{p}=\mu_{o}+\theta (d)\frac{a_o}{\sqrt{d}},
\end{equation}
where $\mu_{o}$ is the contribution from non-critical images, $\theta(d)$ the Heaviside step function, and $a_{o}$ is a constant . This result assumes that locally the caustic is a straight line. It also assumes that the source position remains approximately constant with respect to any other nearby caustics so that the value of $\mu_{o}$ is also constant. In the calculations that follow, Eqn \ref{nearcaust} is used at distances $< 10^{-2}\eta_{o}$, that can be compared to the curvature of the caustics and the inter-caustic distances of $\sim 1\eta_{o}$. Fluke \& Webster (1998) have obtained a higher order approximation to Eqn \ref{nearcaust} that treats the shape of the caustic as a parabola. A possible direction for future work will be to use this result together with information on the caustic curvature to obtain a better approximation to the magnification during both the interpolation and integration algorithms.

To utilise Eqn~\ref{nearcaust} in the evaluation of Eqn~\ref{2dint} we need to know the location of the intersection of the caustic and  the source lines. This is determined during the contouring algorithm, when the approximation Eqn~\ref{nearcaust} with $\mu_{o}=0$ is used for interpolation of magnifications between two source points in the high magnification regime (Lewis et al. 1993). We implement the contouring algorithm as described by Lewis et al (1993). The image solutions are found by stepping along the image curves in the one direction. Therefore the case of the image line crossing a critical curve is determined from a sign change in the magnification corresponding to a parity flip of the image. This sign change results in the direction of travel of the corresponding source line solution being reversed at the caustic, the location of which is then found through a linear interpolation in $\mu_{p}^{-2}$. During the contouring algorithm the intersections of all caustics with each source line are recorded for use in the convolution (Eqn \ref{2dint}).

\begin{figure*}
\begin{center}
\vspace*{65mm}
\includegraphics{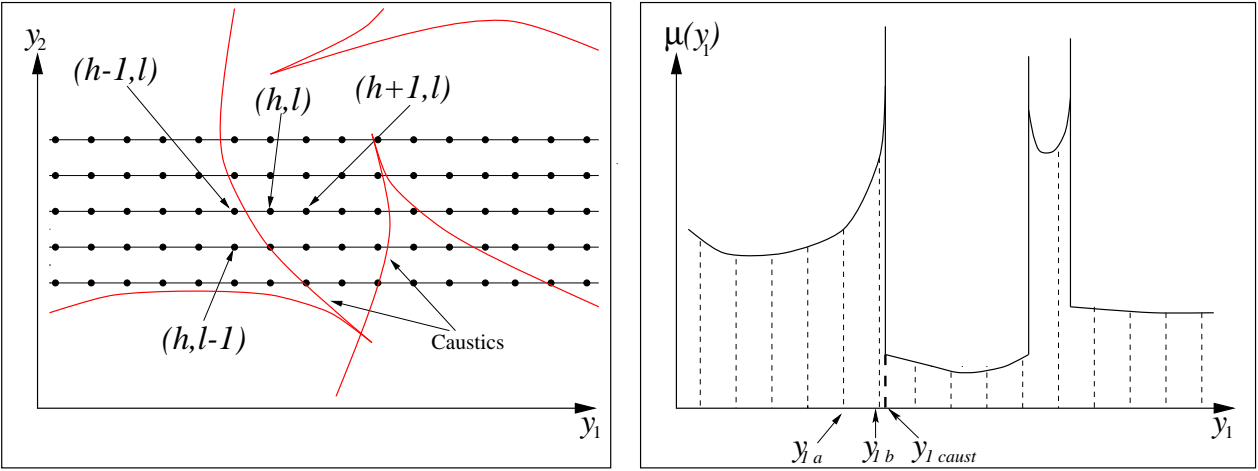}
\caption{ Left: Representation of of the points on a finite source grid, with the labelling as referred to in the text, and Right: Representation of a curve fitted to the amplifications at grid points on the middle source line which crosses the caustic structure shown. The positions of the grid points are labelled by the light dashed lines.}
\label{intschem}
\end{center}
\end{figure*}

Figure \ref{intschem} shows a schematic representation of the source grid in relation to the caustic. The labels of source positions and magnifications which follow refer to this figure and describe their relationship to the source grid.
 The determined positions of the caustics $\vec{y}_{caust}$ are used in conjunction with the values of $\vec{y_{a}}$,$\vec{y_{b}}$ and $\mu_{p}(\vec{y_{a}})$,$\mu_{p}(\vec{y_{b}})$  at the two nearest grid points (inside the caustic), and the straight caustic approximation (Eqn \ref{nearcaust}) to obtain the values of $\mu_{o}$, and $a_{o}$ for the current source point, expressions for which are 
\begin{equation} 
a_{o}=\frac{\mu_{p}(\vec{y}_{a})-\mu_{p}(\vec{y}_{b})}{(|\vec{y}_{caust}-\vec{y}_{a}|)^{-
\frac{1}{2}}-(|\vec{y}_{caust}-\vec{y}_{b}|)^{-\frac{1}{2}}}
\label{2ptint}
\end{equation}
and 
\begin{equation}
\mu_{o}=\mu_{p}(\vec{y}_{a})-\frac{a_{o}}{(|\vec{y}_{caust}-\vec{y}_{a}|)^{\frac{1}{2}}}.
\label{2ptint2}
\end{equation}
Note that the value of $a_{o}$ obtained is not identical to that in Eqn \ref{nearcaust} but is weighted by a factor of $\sqrt{1/sin{\theta}}$ where $\theta$ is the angle between the source trajectory and the tangent to the caustic. This factor cancels with that relating the perpendicular distance $d$ with the distance to the caustic along the source line ($|\vec{y}_{caust}-\vec{y}_{a}|$).
 Since the singularities in $\mu_{p}(\vec{\eta})$ at the caustics are removed in the initial source line integrations, this approach is clearly only valid for integrations along source trajectories that cross caustics having no part within the source strip tangential to the source line. Before integrating out the singularity, the convolution routine must locate the
 two grid points straddling the caustic, and decide which of the two points
 is inside the caustic and which is outside.

 The presence of a caustic between two adjacent points $($h,l$)$ and $($h$\pm$1,l$)$ on a given grid line is determined using the following
 rules; firstly, the magnification $\mu_{p}($h,l$)$ of the current point must be a local maximum
 that is greater than $\mu_{p}=10$, and secondly, the ratio between the point
 source magnifications at the current grid point and that immediately in the forward $\mu_{p}($h$+$1,l$)$
 or backward $\mu_{p}($h$-$1,l$)$ direction must be greater than 10. This second condition ensures that the
 integration of Eqn~\ref{2dint} using the approximation (Eqn \ref{nearcaust}) is not
 carried out in a high magnification but smoothly varying region such as that located
 outside of a cusp. In addition, the second condition determines whether the current source point is immediately inside or immediately outside the caustic, and so which grid points should be used in the evaluation of Eqns \ref{2ptint} and \ref{2ptint2}.
The situation of
 non-isolated caustics is recognised by the algorithm when the difference between
 the position of the current grid point and the recorded position of the
 corresponding caustic ($|y_{1\,a}-y_{1\,caust}|$) is greater than the separation between adjacent grid points ($y_{1\,diff}=|y_{1\,a}-y_{1\,b}|$) on
 the source line. In this case the interpolation scheme (Eqn \ref{nearcaust}) has failed during the contouring algorithm. The routine also looks for another caustic lying along the source
 line within $3$ grid points of the first, and if either of these are the case,
 discounts the contribution to the flux from the point closest to the current caustic since the integration
 of Eqn~\ref{2dint} will not be valid. It should be noted that the routine was
 observed to be robust against variations in the parameters used in these tests.

\subsection{A Comparison Between Light Curves Computed Using the Ray-Shooting and  Contouring Techniques.}

\begin{figure*}
\vspace*{155mm}
\centerline {
\includegraphics{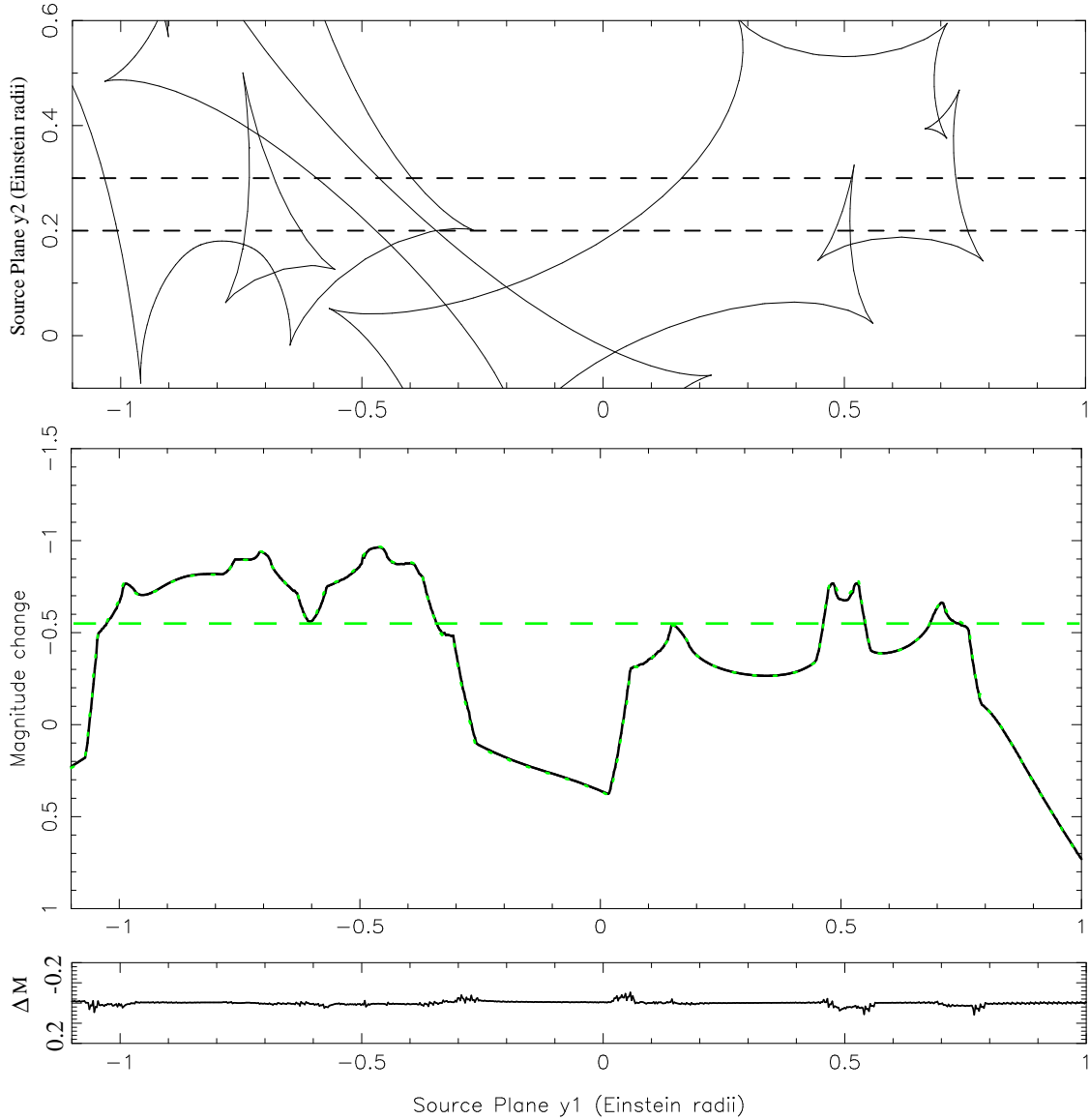}
}
\caption{Middle: The light curve from the accretion disc source calculated using the contouring (solid line) and ray shooting (dotted line) algorithms. The convolution (Eqn~\ref{2dint}) was performed on a $50$ line grid, the source scale is in units of $\eta_{o}$ and the horizontal dashed line is the average magnification for a model with parameters $\kappa=0.2$, $\gamma=-0.2$. Bottom: The difference between the two light curves. Top: The corresponding caustic network. The extremities of the source strip are plotted as dashed lines superimposed on the caustic structure.}
\label{lctest}
\end{figure*}

The backward ray tracing or ray shooting technique of Kayser, Refsdal \& Stabell (1986) provides an excellent test of the validity of a light curve computed using the contouring method. An example of a microlensed light curve for an extended source, calculated using contouring technique (solid line) is shown in Fig~\ref{lctest}. Also shown is the corresponding curve calculated from ray shooting (dotted line), and the numerical difference between the two (lower panel). The corresponding caustic network is shown in the top panel together with the extremities of the source strip and allows features in the light curves to be related to locations on the caustic network. Details of the simulation parameters are described below.

The light curves are calculated for the approximation to the bolometric luminosity of a thin accretion disc with outer radius $R_{s}=0.05\eta_{o}$ (see Grieger, Kayser \& Refsdal 1987). The surface brightness profile is described by the following expression, where $r$ is the radius:
\begin{eqnarray} 
\nonumber P_{accr}(r) &=& \left(\frac{r}{R_{s}}\right)^{-3}\,\,\,\,\,for\,\,\,\,\,\:\frac{R_{s}}{2}\,<\,r\,<\,R_{s}. \\
P_{accr}(r) &=& 0 \hspace{14mm} otherwise
\label{acc}
\end{eqnarray}
\noindent This profile was chosen as a possible
representation of the profile of the continuum region in an active galactic nucleus.  
The source strips were computed with 50 source lines separated by 0.002$\eta_{o}$,
so that the convolution (Eqn \ref{2dint}) for the  contoured curve was computed on a 50$\times$50 grid at each source point $\vec{\eta}$. For the calculation of the ray-shooting source strip, the rays were shot in a regular grid in the lens plane, and the number of rays used was determined so that there were 100 unlensed rays per source strip pixel. The region of lens plane where images were looked for was the same in both the  contouring and ray-shooting calculations. The lens used in this case was 30 point masses with an optical depth of $\kappa\sim$0.2, and a normalised shear of $\gamma=$-0.2 applied parallel to the source line. Calculations of these light curves using source grids composed of $100$ and $200$ lines were also made. Note that this simulation is for comparison between the ray shooting and contouring methods and is not intended as a realistic simulation at the quoted microlensing parameters. However the theoretical average magnification is shown superimposed on the light curves in figure \ref{lctest} (dashed line) for comparison.

 The lower panel of figure~\ref{lctest} shows excellent agreement between the two methods of calculation, particularly at points where the source is not covering the intersection of two caustics or a cusp. In either of these situations, the ray shooting calculation is expected to be more reliable as it is not dependent on the use of the straight caustic approximation. This approximation is not applicable at either the intersection  of two caustics or at source points just inside a cusp where two caustics are within three grid points of each other. In both instances the integration routine will fail since the interpolation scheme is based on this approximation, and so the point source magnifications obtained from the contouring algorithm will be incorrect.  As a quantitative measure of the deviation between the two methods of calculation, the root-mean-squares of the difference between the two light curves during high magnification events on the source trajectory shown in figure~\ref{lctest} were computed for the $50$, $100$ and $200$ line grids. These are listed in table \ref{tab1} and show root-mean-square differences which are of order 1$\%$ of the size of the high magnification event for the calculations using a 50 line grid. The difference between the two methods of calculation is converging to zero for finer grid sizes.

HMEs in microlensed light curves which involve only one caustic exhibit variations in their characteristic shape due to different source geometries (Fluke \& Webster 1998; Schneider, Ehlers \& Falco 1993). This difference between the light curves for the two sources is not a function of the grid size, and we note that it is significantly larger than the numerical difference between the ray shooting and contouring methods. In addition we note that while the numerical uncertainty (ray-shooting minus  contouring) is predominantly present in regions of the light curve where the gradient is largest, the difference in the shape of the light curves is present for the duration of the HME.
\begin{table}
\begin{center}
\caption{\label{tab1} This table refers to the light curves shown in figure \ref{lctest}. The ranges shown describe positions along the source track, and are positioned over the HMEs. The table shows the root-mean-square of the difference between the light curves for the accretion disc source when calculated with the  contouring technique and with the ray shooting technique.}
\begin{tabular}{|c|c|c|c|} \hline
Range of  & rms  & rms  & rms  \\
Source Line $(\frac{\eta_{1}}{\eta_{o}})$ & ($50$ lines) & ($100$ lines) & ($200$ lines) \\ \hline
-1.1 -0.9 & 0.011  & 0.004 & 0.002 \\
-0.8 -0.6 & 0.009  & 0.005 & 0.003 \\
-0.6 -0.2 & 0.014  & 0.008 & 0.005 \\
0.0 0.2 & 0.014    & 0.007 & 0.004  \\
0.4 0.6 & 0.029    & 0.018 & 0.007 \\ 
0.65 0.8 & 0.013   & 0.007 & 0.004  \\ \hline
\end{tabular}
\end{center}
\end{table}

The contouring algorithm has two major advantages over the ray shooting method. Firstly, it solves the lens mapping function, rather than finding the magnifications through a statistical calculation. Secondly, and more importantly, there is an increase in speed of computation. For calculations of light curves from different source geometries, the magnification grid in the source plane must have a small resolution so that the convolution (Eqn \ref{2dint}) can be computed over many grid points.
To obtain an accurate magnification in a pixel using the ray shooting algorithm requires at least 100 un-lensed rays per pixel (Schneider, Ehlers \& Falco 1993). Computation of a fine grid over a long source line therefore becomes impractical for large star-fields and small source sizes. Computation times on a Unix workstation (Sun UltraSPARC 1) for the 50, 100, and 200 line source grids used to calculate the light curves in figure \ref{lctest} were found to be $\sim 50$ seconds, $\sim 100$ seconds and $\sim 200$ seconds for the contouring algorithm, compared with $\sim 150$ minutes, $\sim 600$ minutes and $\sim 2400$ minutes respectively for the same calculation using ray shooting with the deflection of rays being calculated across the minimum region required to collect the total flux (for this lens configuration). During realistic modelling the addition of a large number of point masses means that this region must be enlarged. This is discussed further in section \ref{Simulations1}. The difference in required CPU time is due to the number of times that each algorithm must compute the bending angle. The length of time required for the calculation of a given light curve is approximately independent of source size for contouring but is proportional to the inverse of the square of the source diameter for ray shooting. In addition, computation time increases linearly with the number of source lines for  contouring, but with the square of the number of source lines for ray shooting. The ray shooting method can easily produce a number of light curves from the same model starfield. However, the source tracks will have significantly overlapping shooting regions and will therefore not be statistically independent. Thus, relatively speaking, contouring becomes more efficient for smaller sources and finer grids.

Haugan (1996) suggested the use of contouring to define the regions of the source plane in which rays need to be traced so that they are only collected by the region of interest in the source plane, thus reducing the number of rays required. He obtains an expression for the efficiency of the hybrid method with respect to traditional ray shooting in terms of the number of rays used (not including the contouring calculations). For a simulation of optical depth $\kappa=0.2$ this expression yields an efficiency of $\sim200$ for the source track we have computed for this section. This number is comparable to the efficiency we obtained from computation times for the 50 line strip. We note however that as the hybrid method still employs the ray shooting technique, its efficiency with respect to contouring is still proportional to the inverse of the number of lines in the strip, as well as to the source size. Thus contouring will still be more efficient for smaller sources and finer grids. 

\section{Model Light Curves For Q2237+0305}
\label{Simulations}

The substantial increase in computational efficiency allows a range
of microlensing simulations to be explored. This enables the determination of statistical measures for comparison with the data. 
 
\subsection{Models with Static Stellar Positions}
\label{Simulations1}

\begin{figure*}
\vspace*{90mm}
\centerline {
\includegraphics{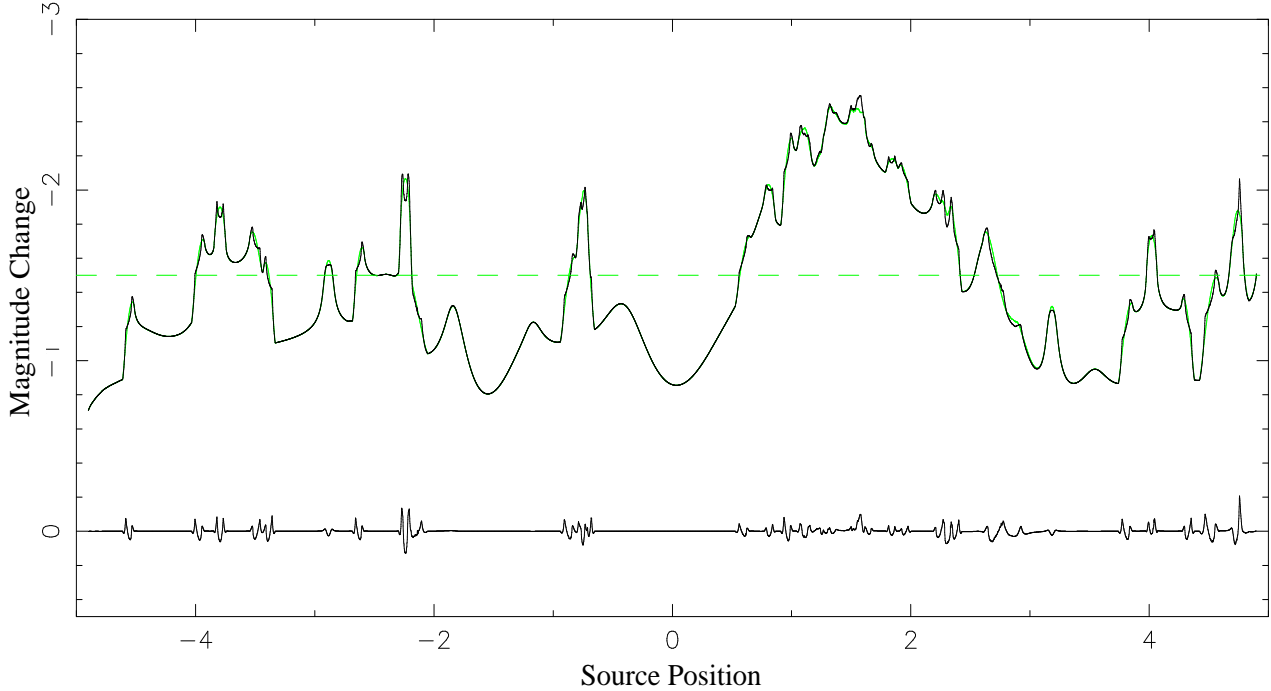}
}
\caption{Sample light curves for the approximation to an accretion disc source (dark line) and a circular source with limb darkening (light line), together with the difference between the two curves. The microlensing parameters are $\kappa=0.36$ and $\gamma=+0.4$.}
\label{sample}
\end{figure*}

The observation of microlensing in Q2237+0305 has spawned many attempts to characterise the variation in flux using numerical microlensing models. The ray shooting method has been used to produce caustic maps and light curves from various source sizes of each of the images in Q2237+0305. Examples of these are given in Wambsganss, Paczynski \& Katz (1989). We base our models upon those presented in that paper. The models consist of point masses with positions distributed randomly in a disc, and masses determined from a Salpeter mass function $p(m)\propto m^{-2.35}$ in the range $0.1M_{\odot}<m_{i}<1.0M_{\odot}$.  The model does not include a continuous matter distribution.  The region of the lens plane in which image solutions need to be found to ensure that $99\%$ of the total macro-image flux is recovered from all points on the source line is known as the shooting region. This region is found by first calculating the image of the source line, as lensed by a smooth screen of matter with the same optical depth as the starfield. The number of stars in the region about any point which collects 99\% of the macro-image flux was calculated by  Katz, Balbus \& Paczynski (1986) and is given by
\begin{equation}
N_{*}=100\frac{\langle m^{2}\rangle}{\langle m \rangle ^{2}}\frac{\kappa^{2}}{|(1-\kappa)^{2}-\gamma^{2}|}.
\end{equation}
In the case of an applied shear these stars are distributed in an ellipse with a major to minor axis ratio of $(1-\kappa+\gamma)/(1-\kappa-\gamma)$. The shooting region is then given by the union of these ellipses centred on all parts of the image line. The minimum number of stars required in the model are contained in a disc having a radius $R_{*}$ which covers this shooting region. In the cases of $\gamma=-0.4$ and $\gamma=+0.4$ we have chosen the radius of the disc of point masses to be $1.2\times R_{*}$ and $R_{*}$ respectively.

Initially, the stars are assumed to be stationary. The effect of including stellar proper motions in the microlensing model is discussed in section \ref{Propermot}. The macrolens model of Schmidt, Webster \& Lewis (1998) was used to determine the microlensing model, giving the parameters of $\kappa=0.36$ and $|\gamma|=0.40$ as a description of the microlensing environment of image A. A transverse velocity in the lens plane of $600 \kms$ was assumed (eg. Wambsganss, Paczynski \& Schneider 1990). The size of the simulations was then determined such that the length of each was $\sim 70$ years for the model with positive shear, and $\sim 140$ years for the model with negative shear. Therefore the length of each of our simulations is $10\eta_{o}$ where $\gamma=+0.4$ and $20\eta_{o}$ where $\gamma=-0.4$. The numbers of stars in our models are then 2032 in the case where $\gamma=+0.4$ and 880 where $\gamma=-0.4$.

For Q2237+0305, the Einstein radius of a $1M_{\odot}$ star in the lensing galaxy  ($z=0.0394$) projected to the quasar redshift ($z=1.695$) is $\eta_{o}\approx1\times 10^{17}\cm$. Light curves were calculated for the cases of $\gamma=-0.4$ and $\gamma=+0.4$ that correspond to source trajectories directed towards, and at right angles to, the centre of the lensing galaxy respectively.  Contouring simulations were calculated  in each of these cases for source radii of $5\times10^{15}\cm\,(R_{s}\sim\frac{\eta_{o}}{20})$ and $1\times10^{15}\cm\,(R_{s}\sim\frac{\eta_{o}}{100})$ using as the source profile an accretion disc model of the source (Eqn \ref{acc}), and an approximation to a circular source with limb darkening (Grieger, Kayser \& Refsdal 1987),
\begin{eqnarray} 
\nonumber P_{cir}(r)&=&\sqrt{1-\left(\frac{r}{R_{s}}\right)^{2}}\,\,\,\,\,for\,\,\,\,\,\:0\,<\,r\,<R_{s}.\\
P_{cir}(r) &=& 0 \hspace{21mm} otherwise
\end{eqnarray}
The two source geometries have different surface areas but are normalised to have the same total unlensed flux.
 Again, the source grids were composed of 50 lines each. The light curves therefore contain 5000, and 25000 points for the $R_{s}\sim\frac{\eta_{o}}{20}$ and $R_{s}\sim\frac{\eta_{o}}{100}$ respectively in simulations with $\gamma=+0.4$, and twice this number in simulations where $\gamma=-0.4$. This was considered a sufficient resolution to determine the presence of single caustic high magnification events, and other miscellaneous events due to clusters of fold caustics.  For comparison of the two sources in a realistic model, examples of the light curves for both the $R_{s}=\frac{\eta_{o}}{20}$ accretion disc (dark line) and circular sources (light line) for the case of $\gamma=+0.4$ are shown in figure \ref{sample}. The difference between the light curves for the circular and accretion disc profiles is also plotted, and shows that differences are present in the model up to a level of 0.1 magnitudes. 

A random starfield will be homogeneous when considered on a large enough scale so that the granular nature of its boundary will have a negligible effect. Wambsganss, Paczynski \& Katz (1989), and others have used tree-codes and associated methods to calculate the bending angle due to fields of as many as $10^{5}$ point masses. These codes have the effect of smoothing the matter far from the image so that only the contributions to the bending angle from the few hundred point masses nearest to the image position are summed directly. The magnification patterns produced in microlensing simulations respond in a complicated, and often radical manner to small changes in the position or mass of one of the model stars (Wambsganss 1992). However, this change only occurs in parts of the source plane where the corresponding shooting region includes the point mass in question. A model starfield is therefore sufficiently large if it covers the shooting regions corresponding to all points on the source line. In the case of the models used in the current work, the simulations were sufficiently small that we found the use of tree codes to be unnecessary.

\begin{figure}
\vspace*{65mm}
\centerline {
\includegraphics{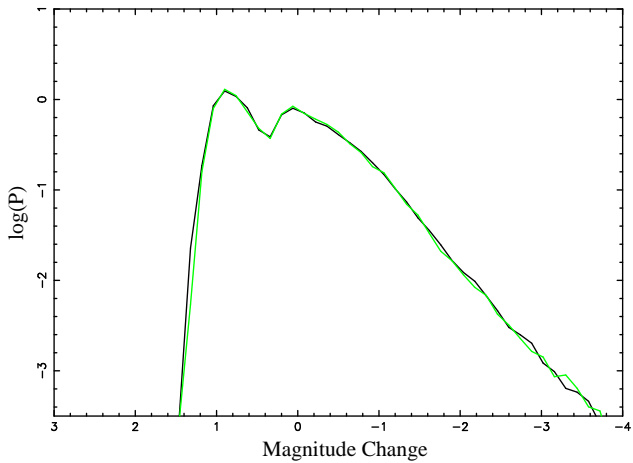}
}
\caption{Point source magnification distributions for the parameters $\kappa=0.36$ and $\gamma=+0.4$ (dark line), $\gamma=-0.4$ (light line).}
\label{amphist}
\end{figure}

To check that the parameters used in our simulations are appropriate we generated a sample of 100 point source light curves, in each of the cases $\gamma\pm0.4$ using the source line lengths and parameters described previously . We find that that the mean magnifications for the $\gamma>0$ and $\gamma<0$ light curves are 3.92 and 3.90 respectively. These values are consistent with the theoretical value of $\mu_{av}=1/((1-\kappa)^{2}-\gamma^{2})=4.00$. In addition, the difference between theoretical and numerical values is consistent with that quoted for comparable models in (Lewis \& Irwin 1995). This demonstrates that our models are indeed finding the micro images that account for the macro image flux. Figure \ref{amphist} shows the probability that a point source in our models is magnified by a magnitude between $m$ and $m+\delta m$, where $m=-2.5\times log_{10}(\mu /\mu_{av})$ and $\mu$ is the microlensing magnification generated from our models in the cases of $\gamma>0$ (dark line) and $\gamma<0$ (light line). As required these curves are identical. In addition we can compare the distributions to those in Lewis \& Irwin (1995). Our distributions are quantitatively consistent with those presented in that paper.      

\subsection{High Magnification Events}

\begin{table*}
\begin{center}
\caption{\label{simlevents} The frequency of different classes of HME in the various classes of simulation. All measurements of time assume a lens plane transverse velocity of 600$\, km \, sec^{-1}$.}
\begin{tabular}{|c|c|c|c|c|c|c|} \hline
Class of   & Total length of & Cusp HMEs & SHMEs & MHMEs &total HMEs & fraction\\
Simulations& simulations (years) &  per decade & per decade & per decade & per decade & of SHMEs \\ \hline
$R_{s}=.05\eta_{o}\;\gamma =+0.4$  & 1610 &$ 0.12 \pm .02$ &$ 1.21 \pm .07$ & $0.87 \pm .06 $ & $ 2.20 \pm .10 $ & 0.55 $\pm$ .04 \\
$R_{s}=.01\eta_{o}\;\gamma =+0.4$  & 420 & $ 0.18 \pm .05$ &$ 2.47 \pm .20$ & $ 0.38 \pm .08 $ & $ 3.03 \pm .22 $ & 0.82 $\pm$ .09 \\
$R_{s}=.05\eta_{o}\;\gamma =-0.4$  & 1680 &$0.13 \pm .02$ &$ 0.63 \pm .05$ & $ 0.62 \pm .05 $ & $ 1.38 \pm .08 $ & 0.46 $\pm$ .05 \\
$R_{s}=.01\eta_{o}\;\gamma =-0.4$  & 700 & $0.16 \pm .04 $& $1.16 \pm .10$ & $ 0.19 \pm .04 $ & $1.51 \pm .12 $ & 0.77 $\pm$ .09 \\ \hline
\end{tabular}
\end{center}
\end{table*}

In order to analyse the high magnification events (HMEs), three different
classes are  defined.
\begin{enumerate}
\item Single-Caustic High Magnification Events (SMHE): HMEs that have the same general shape as the corresponding convolution of the profile with the straight caustic approximation (see for example Schneider, Ehlers \& Falco (1993)). The event must be complete, having a discontinuity in  gradient at its leading edge as well as a drop in magnitude at its trailing end. It must include the two discontinuities in gradient, the small secondary peak, and the 4 changes in the sign of the second derivative of the curve. 
\item Multi-Caustic High Magnification Events (MHME): HMEs that have both a rise and a fall, as well as at least one discontinuity in their gradient. These are not recognisable as SHMEs as the source straddles two or more fold caustics. 
\item Cusp Events: HMEs where the source moves through a high magnification region outside a cusp. They have both a rise and a fall but no discontinuities in their gradient and are conspicuous in the model light curves because the shape of the HME is approximately independent of the source geometry, and approximately symmetric about its maximum.
\end{enumerate}
The HMEs were classified interactively.

 Table~\ref{simlevents} shows the average rates for the different types of HME in each of the four cases described earlier. These rates depend linearly on the transverse velocity. The errors in these, and results of later sections, were calculated as the Poisson error of the total number of events in the light curves. As was noted by Witt, Kayser \& Refsdal (1993), and others, the rate of variability is greater for a source which has a trajectory parallel to the shear vector ($\gamma>0$) due to the stretching of the caustic network. This change in rate is due to the fact that a source moving perpendicularly to the shear vector ($\gamma<0$) will on average spend more time in regions of the source plane which contain no caustic clusters. For image A of Q2237+0305, and a small source the simulations of the current work show this change in rate to be a factor of approximately 2, in agreement with the results of Lewis \& Irwin (1996). 

Since the direction of the source trajectory has a significant effect on the frequency of observed HMEs, there is a degeneracy between the effective transverse speed of the source and the direction of the source trajectory. In the case of Q2237+0305, it may be possible to eventually deduce the direction of  motion relative to the caustic network by continued monitoring of the four images, which are orthogonally positioned with respect to the centre of the lensing galaxy (eg. Witt \& Mao 1994). However, the simulations show that we also expect that the ratio of the frequency of Cusp HMEs to total HMEs to be approximately twice as large for a source moving along a path perpendicular to the shear vector than parallel to it. Since this ratio is independent of the transverse speed, its measurement provides a direct method of determining the direction of motion. In addition this ratio appears to be independent of source size. The higher proportion of cusp events in the case of negative shear is expected because the preferential direction of formation of the caustics with respect to shear vector directly implies that the cusps also have a preferred direction. Isolated cusps form mostly on the outside of the long caustic clusters so that a source trajectory that is parallel to the direction of the caustic clustering will encounter more of these smooth high magnification regions. 
    
The total frequency of HMEs is a function of source size, with a higher frequency for smaller sources. This increase is not large for the range of source sizes considered here, and is due to the fact that a smaller source is less likely than a larger source to cover multiple caustics. In contrast, the relative proportion of SHMEs and MHMEs is a very sensitive function of source size, with SHMEs being $\sim$ 6 times as prevalent (with respect to MHMEs) for the $R_{s}=\frac{\eta_{o}}{100}$ source than for the $R_{s}=\frac{\eta_{o}}{20}$ source. This result appears to be approximately independent of trajectory direction.

In order to determine the statistical frequency of HMEs, frequent monitoring for a long period of time is required. The most important scientific result is the large percentage of HMEs in the model light curves that are also SHMEs. If the central region of the source quasar has a geometrically simple intensity profile then we would expect to see a characteristic shape in the observed light curves. This characteristic event shape could then be deconvolved to reconstruct the source intensity profile either by trial and error, or by a reconstruction method such as that suggested by Grieger, Kayser \& Refsdal (1987). Information on the source intensity profile might be obtained after  observing just a few HMEs.

\subsection{The Effect Of Proper Motion Of Stars }
\label{Propermot}

Most of the numerical work on microlensing in Q2237+0305 has concentrated on models of static starfields, with the flux variation resulting from an assumed effective transverse velocity in the lens plane of $600 \kms$. The main motivation for this assumption is undoubtedly the enormous increase in computation time required to compute a light curve with the inclusion of stellar proper motions. This increase is the result of the entire magnification pattern having to be re-computed for each point on the model light curve. In addition to the positions of the point masses, the macro parameters are in principle also time dependent. However, these time variations are negligible when compared with the motion of stars because changes in $\kappa$ and $\gamma$ occur over a scale of more than 1000 years (Kundic, Witt \& Chang 1993).

\begin{table*}
\begin{center}
\caption{\label{movsimlevents} The frequency of different HMEs in microlensing models that include stellar proper motions. All measurements of time assume a lens plane transverse velocity of 600$\, km \, sec^{-1}$.}
\begin{tabular}{|c|c|c|c|c|c|} \hline
Class of   & Total length of & Cusp HMEs in & SHMEs in & Total HMEs & fraction \\
Simulations & simulations (years) & both curves & both curves & per decade & of SHMEs\\
            &                     & per decade & per decade  & \\ \hline
$R_{s}=.05\eta_{o}\;\gamma =+0.4\;$  & 770 & $ 0.09\pm .03 $&$ 0.89\pm .09 $& $2.26\pm .15$ & 0.39 $\pm$ .05\\
$R_{s}=.01\eta_{o}\;\gamma =+0.4\;$  & 350 & $0.24\pm .06$ & $2.18\pm .20 $& $3.06\pm .24$ & 0.71 $\pm$ .09\\
$R_{s}=.05\eta_{o}\;\gamma =-0.4\;$  & 1260 & $0.09\pm .03 $& $0.49\pm .06$ & $1.43\pm .09$ & 0.34 $\pm$ .05\\
$R_{s}=.01\eta_{o}\;\gamma =-0.4\;$  & 700 & $0.19\pm .05 $& $0.94\pm .11 $& $1.44\pm .14$ & 0.65 $\pm$ .10\\ \hline
\end{tabular}
\end{center}
\end{table*}

The observed one dimensional velocity dispersion of the bulge in the galaxy of Q2237+0305 is $\approx 215\, \kms$ (Foltz et al. 1992). If this value is representative of the dispersion at the position of image A, then it may have a significant impact on the statistics of HMEs. Two approaches have been taken towards the computation of the effect of a velocity dispersion on microlensing statistics. The velocity of the caustic network resulting from stellar velocities has been calculated by Schramm et al. (1993), and by Kundic, Witt \& Chang (1993). The latter consider the area swept out per unit time by the caustic network produced from an ensemble of point masses with a velocity dispersion. They find that a one dimensional velocity dispersion will produce more HMEs than a static model with the equivalent transverse velocity. A second approach has been taken by Kundic \& Wambsganss (1993) and Wambsganss \& Kundic (1995). They investigate the microlensing contribution by computing the change in the magnification pattern that occurs when the model stars are ascribed motions according to a velocity dispersion.  Kundic \& Wambsganss (1993) find a contribution to HMEs from the velocity dispersion which is consistent with that of Kundic et al. (1993). For image A in Q2273+0305 Kundic \& Wambsganss (1993) find that a one dimensional velocity dispersion of $\sigma\approx 215\,km\,sec^{-1}$ will increase the number of observed HMEs by about 10\% if the transverse velocity is assumed to be $v_{t}=600\,km\,sec^{-1}$.

 Since stellar proper motions may have a large effect on the statistics of HMEs, it is necessary to investigate their effect on the existence and relative proportions of the different types of HMEs. Our prime motivation is to estimate the expected prevalence of characteristic SHMEs in an observed light curve. Therefore, rather than compute many points along a light curve, each with a slightly evolved starfield, many of the simulations from section \ref{Simulations1} were re-run with the stars having been displaced in both the $y_{1}$ and $y_{2}$ directions by an amount determined from a speed distributed according to a Gaussian having a dispersion of $215\, \kms$, (note this assumes an isotropic velocity distribution), and a time equal to that required for a caustic with the projected bulk velocity to move $1\frac{1}{2}\times$ the source width. If a characteristic SHME is present in both the light curves, then it will also be present in a fully computed curve using a model with an evolving field. The time scale of this event would be either increased or decreased depending on whether the fold caustic was moving with or against the bulk velocity. The proportion of total HMEs that are SHMEs obtained in this manner will approximate the ratio SHME:HME for the model used.

Table~\ref{movsimlevents} gives the relative frequency of the different classes of HMEs when stellar proper motions are included in the microlensing model. These rates are obtained using the effective transverse velocity and do not take into account the increased frequency of HMEs due to the inclusion of stellar proper motions. The incorporation of proper motion has had the effect of reducing the proportion of SHMEs in the model light curves. This is a consequence of the evolving caustic network, however the frequency of SHMEs is still an appreciable fraction of that obtained in section~\ref{Simulations1}.  Therefore the effect of proper motion of stars in the model is to increase the overall number of HMEs, but to decrease the fraction of SHMEs.

Whilst the effective transverse velocity is unknown, and may well be relatively small, the fact that stellar proper motions can produce microlensing events at an appreciable rate, means that even in the absence of a transverse velocity, both HMEs and SHMEs are common. The presence of a SHME in one of the time separated light curves but not the other is due to that event having been corrupted by a second, or several more fold caustics moving into the source path as the caustic structure evolves. The drop in the proportion of total HMEs which are SHMEs in both of the time separated light curves relative to the proportion obtained using a static field is, as it must be, offset by an increase in the frequency of MHMEs. The proportion of cusp HMEs in the model light curve has been unchanged by the inclusion of proper motion. This is expected because these events are formed when the source moves through a high magnification region outside the parts of the source plane which are densely populated with caustics.

\section{Conclusion}

A new method has been developed to calculate the microlensed light curve of an extended source. In realistic simulations this method achieves a given accuracy for a light curve in less than one tenth of a percent of the time required by the ray-shooting method. The gain in efficiency has been used to obtain extra resolution in realistic model light curves. This has allowed an investigation of different types of high magnification events to be made. 

Long term monitoring provides a method for measuring 
the physical parameters of the source and lens.  As an example, the ratio of cusp to total HMEs is approximately twice as large for a source moving perpendicular to the shear than for one moving parallel to the shear. This result is independent of source size and could be used to obtain information on the direction of the source trajectory, a variable which is required in the determination of the  transverse velocity. In addition, we note that SHMEs will be far more common with respect to MHMEs for smaller source sizes. This result is approximately independent of trajectory direction and so could potentially be used to determine the value of the source size.

Between 50\% and 75\% of the HMEs were found to be SHMEs in the static field model light curves. The introduction of stellar proper motions into the microlensing model had the effect of lowering this fraction.  However the drop was not significant -- less than 20\%. The relatively high fraction of SHMEs is a very exciting result since these events will allow the geometry
of the continuum region of the source quasar to be measured from a single well sampled HME.

Q2237+0305 has three images other than the one modelled here. Image B has parameters similar to those of image A, while images C and D are located in regions which have a higher optical depth. One direction for future work will be to repeat this analysis for model star fields which have optical depths corresponding to these other images.

\label{lastpage}


\begin{thebibliography}{}

\bibitem[\protect\citename{Chang \& Refsdal }1979]{CH79}
Chang, K., Refsdal, S., 1979, Nature, 282, 561 

\bibitem[\protect\citename{Chang }1984]{CH84}
Chang, K., 1984,  Astron. Astroph., 132, 168 

\bibitem[\protect\citename{Corrigan \etal }1991]{CO91}
Corrigan et al., 1991, Astron. J., 102, 34

\bibitem[\protect\citename{Fluke \& Webster }1998]{FL98}
Fluke, C. J., Webster, R. L.,  1999, M.N.R.A.S., 302, 68

\bibitem[\protect\citename{Foltz \etal  }1992]{FO92}
Foltz, C. B., Hewitt, P. C., Webster, R. L., Lewis, G. F., 1992, Ap. J., 386, L43

\bibitem[\protect\citename{Grieger, Kayser \& Refsdal }1987]{GR87}
Grieger, B., Kayser, R., and Refsdal, S., 1987,  Astron. Astrophys., 194, 54

\bibitem[\protect\citename{Haugan }1996]{HA96}
Haugan, S. V. H., 1996, IAU Sypmosium 173, p275.

\bibitem[\protect\citename{Irwin \etal  }1989]{IR89}
Irwin, M. J., Webster, R. L., Hewitt, P. C., Corrigan, R. T., Jedrzejewski, R. I., 1989, Astron. J., 98, 1989

\bibitem[\protect\citename{Katz, Balbus \& Paczynski }1986]{KA86}
Katz, N., Balbus, S., Paczynski, B., 1986, Ap. J., 306, 2

\bibitem[\protect\citename{Kayser, Refsdal \& Stabell }1986]{KA86b}
Kayser, R., Refsdal, S., Stabell, R., 1986, Astron. Astrophys., 166, 36

\bibitem[\protect\citename{Kundic \& Wambsganss }1993]{KU93}
Kundic, T., Wambsganss, J., 1993, Ap. J., 404, 455

\bibitem[\protect\citename{Kundic, Witt \& Chang }1993]{KU93b}
Kundic, T., Witt, H. J., Chang, K., 1993, Ap. J., 409, 537
 
\bibitem[\protect\citename{Lewis \etal   }1993]{LE93} Lewis, G. F., Miralda-Escude, J., Richardson, D. C., Wambsganss, J., 1993, MNRAS, 261, 647

\bibitem[\protect\citename{Lewis \& Irwin }1996]{LW96}
Lewis, G. F., Irwin, M. J., 1995, MNRAS, 276, 103

\bibitem[\protect\citename{Lewis \& Irwin }1996]{LW96}
Lewis, G. F., Irwin, M. J., 1996, MNRAS, 283, 225

\bibitem[\protect\citename{Paczynski }1986]{PA86}
Paczynski, B., 1986, Ap. J, 301, 503


\bibitem[\protect\citename{Schmidt, Webster \& Lewis }1998]{SC98}
Schmidt, W. R., Webster, R. L., Lewis, G. F. 1998, MNRAS, 295, 488

\bibitem[\protect\citename{Schmidt \& Wambsganss }1998]{SC98b}
Schmidt, W. R., Wambsganss, J., 1998, Astron. Astrophys., submitted

\bibitem[\protect\citename{Schneider, Ehlers \& Falco }1993]{SC93}
Schneider, P., Ehlers, J., Falco, E.E., 1993, Gravitational lenses, Springer-Verlag, Germany

\bibitem[\protect\citename{Schramm et al.}1993]{SC93}
Schramm, T., Kayser, R., Chang, K, Nieser, L., Refsdal, S., 1993, Astron. Astrophys., 268, 350

\bibitem[\protect\citename{Wambsganss \& Kundic }1995]{WA95}
Wambsganss, Kundic, T., 1995, Ap. J., 450, 19

\bibitem[\protect\citename{Wambsganss, Witt, \& Schneider}1995]{WK95}
Wambsganss, J., Witt H., J., Schneider, P., 1992, Astron. Astrophys., 258, 591


\bibitem[\protect\citename{Wambsganss, Paczynski \& Katz}1990]{WA90}
Wambsganss, J., Paczynski, B., Katz K., 1990, Ap. J., 352, 407

\bibitem[\protect\citename{Wambsganss, Paczynski \& Schneider }1990]{WA90b}
Wambsganss, J., Paczynski, B., Schneider, P., 1990, Ap. J., 358, L33

\bibitem[\protect\citename{Wambsganss \& Paczynski }1991]{WA91}
Wambsganss, J., Paczynski, B., 1991, Astron. J., 102, 864
 
\bibitem[\protect\citename{Wambsganss }1992]{WA92b}
Wambsganss, 1992, Ap. J., 392, 424

\bibitem[\protect\citename{Witt }1993]{WI93}
Witt, H. J., 1993, Ap. J., 403, 530

\bibitem[\protect\citename{Witt et al.}1993]{WI93_2}
Witt, H. J., Kayser R., Refsdal, S., 1993, Astron. Astrophys., 268, 501

\bibitem[\protect\citename{Witt \& Mao }1994]{WI94}
Witt, H. J., Mao, S., 1994, Ap. J., 429, 66

\end{thebibliography}
\end{document}